\documentstyle[aps,prd,preprint]{revtex}
\newcommand{\be}{\begin{equation}}
\newcommand{\ee}{\end{equation}}
\newcommand{\ba}{\begin{eqnarray}}
\newcommand{\ea}{\end{eqnarray}}
\newcommand{\bm}{\bibitem}

\begin{document}

\title{Entropy of Scalar Field in 3+1 Dimensional 
Reissner-Nordstrom de Sitter
Black Hole Background}
\author{K. Ghosh\footnote{E-address: kaushik@theory.saha.ernet.in}} 
\address{ Saha Institute of Nuclear Physics, 1/AF, Bidhannagar,
Calcutta - 700064, India.}

\maketitle
\vspace{1cm}

\begin{abstract}

We consider the thermodynamics of minimally coupled massive
scalar field in 3+1 dimensional Reissner-Nordstrom de Sitter
black hole
background. The brick wall model of 't Hooft is used. When
Schwarzschild like coordinates are used it is found 
that two radial brick wall cut-off parameters are required
to regularize
the solution. Free energy of the scalar field is 
obtained through counting of states using the WKB
approximation. It is found that the free energy 
and the entropy are divergent in both
the cut-off parameters.

\end{abstract}

\newpage

\section{{Introduction}}

A black hole has a horizon beyond which
no matter or information can escape. The
absence of information about the region inside the horizon
manifests itself in an entropy. A quantitative expression
for the entropy and the laws of black hole thermodynamics
were first obtained by Bekenstein [1] mainly on the basis of analogy.
Since then a lot of effort has been devoted to explain this entropy on
a statistical mechanics basis. A related issue
is the entropy of quantum fields in black hole backgrounds.
The  entropy  of  quantum  fields is obtained by various methods,
e.g.,
by tracing over local degrees of freedom inside the horizon
(geometric entropy) [2], by explicit counting of degrees of freedom
of the fields propagating outside the horizon
(entanglement entropy) [3,4,5] or by 
the Euclidean path integral [6,7].
These expressions are proportional to the area of the horizon
and constitute the first quantum correction to the gravitational
entropy. Divergences appear in the density of
the states associated to the horizon and can be absorbed in the
renormalized expression of the gravitational coupling constant [8].
To regularize these divergences 't Hooft proposed [3] that the field modes
should be cut off in the vicinity of the horizon by imposing
a brick wall cut-off. This method has  been  used  to  study  the
entropy of matter around different black hole solutions.

All recently available data from cosmological observations,
including measurements of the present value of the Hubble
parameter and dynamical estimates of the present energy
density of the Universe, give strong sugessions that
in the framework of inflationary cosmology a nonzero
repulsive cosmological constant, 
$\Lambda \ge 0,~(10^{16}-10^{18}~GeV)^2$,
has to be invoked in order to explain the properties
of the presently observed Universe.
De Sitter space is the maximally symmetric solution
of the vacuum Einstein equations with a positive
cosmological constant $\Lambda$.
Ginsparg and Perry [9] studied the semiclassical
instabilities of de Sitter space. Due to the presence
of a cosmological event horizon and its associated
Hawking radiation, the de Sitter space exhibits a
semiclassical instability to the nucleation of
black holes, known as the Kottler [10] or Schwarzschild
-de Sitter metric. This solution represents a
nonrotating black hole immersed in de Sitter space.   
Mann and Ross [11] studied the pair creation of
electrically charged black holes in a cosmological
background. There are two instantons describing the
pair production of non-extreme and extreme black holes
respectively. These solutions represents nonrotating
electrically charged black holes immersed in de Sitter
space and are called the Reissner-Nordstrom de Sitter
black holes.

Hawking and Gibbons [12] studied the thermodynamical 
aspects and the Hawking radiation  of scalar fields
in the Kerr-Newman-de Sitter space. It can be shown
that if we try to use the Euclidean Schwarzschild-
de Sitter solution to provide thermodynamical 
aspects, the time periods required to avoid the
conical singularities at the black hole horizon
and the cosmological horizon, do not match. This
is physically interpreted as indicating that the
two horizons are not in thermal equilibrium and
they both emit Hawking radiation at the corresponding
temperatures. An observer situated somewhere between
the black hole horizon and the cosmological horizon
and whose world line coincides with an orbit of
the static killing vector  receives a thermal
radiation coming from the black hole and an isotropic
thermal radiation with a different temperature coming  
from the cosmological horizon. However, in the case of
Reissner-Nordstrom de Sitter
black holes it has been observed that there are two
sectors of solutions depending on the values of
the parameters present in the metric for which
the black hole horizon and the cosmological horizon
are in thermal  equilibrium. These are known as the 
lukewarm Reissner-Nordstrom de Sitter black hole and
the cold Reissner-Nordstrom de Sitter black hole
solutions.
  
In this context it is natural to enquire about
the entropy of quantum fields defined on such backgrounds.
Thus we investigate the thermodynamical behavior of a 
massive minimally coupled real
scalar field propagating on the 3+1 dimensional
lukewarm Reissner-Nordstrom de Sitter black hole and
the cold Reissner-Nordstrom de Sitter black hole
solutions using the brick wall cut-off.

\section{{Renormalization
of the gravitational action and entanglement entropy}}

In the study of the one-loop effective action [13],
we may start with the gravitational action
\be
I_{g} = {{\int}{d^4 x}}{\sqrt{-g}}{\big[{{-{\Lambda\over{8\pi G}}}
+ {R\over{16\pi G}}}\big]}
\ee
where $\Lambda$ is the cosmological constant and $G$
is Newton's constant. Here we have neglected the
interactions which are quadratic in the curvature.
The constants are bare coupling constants.
In the case of Reissner-Nordstrom black hole there will
be a Maxwell term and, in general, additional higher
derivative interactions with the metric and gauge fields:
\be
I_{M} = {\int}{d^4 x}{\sqrt{-g}}{\big[{{-{1\over4}{F_{ab}}{F^{ab}}}
+ {{\delta}({F_{ab}}{F^{ab}})^2} + {\lambda{R_{ab}}{F^{ac}}{F_c ^b}}}\big]}
\ee
here $\delta$ and $\lambda$ are the bare coupling constants. We also
include the action for a minimally coupled neutral scalar field,
\be
I_{m} = {\int}{d^4 x}{\sqrt{-g}}{\big[{g^{ab}{\bigtriangledown}_{a}{\psi}
{\bigtriangledown}_{b}{\psi} + m^{2}{\psi}^{2}}\big]}
\ee 
We want to study the effective action for the metric which results
when in the path integral the scalar field is integrated out. In 
this case , the integration is simply gausian, yielding the
square root of the determinant of the propagator. The contribution
to the effective gravitational action is given by,
\be
W(g) = -{i\over 2}Tr log[-G_F(g,m^2)]
\ee
This expresssion is divergent. The divergence of this one-loop
effective action and its metric dependence can be found through
an adiabatic expansion for the DeWitt-Schwinger proper time
representation of the propagator [14]. This leads to a 
representation of the scalar one-loop action as an asymptotic
series [14]:
\be
W(g) = -{1\over 32\pi^{2}} {\int}{d^4 x}{\sqrt{-g}}
{{\int}_{0}^{\infty}}{ds\over s^3}{\Sigma}{a_n (x)}{(is)^n}{e^{-im^2 s}}   
\ee
here the summation runs over all integral values of $n$ from
$0$ to $\infty$. The adiabatic expansion coefficients $a_n (x)$
are functionals of the local geometry at $x$ and can be constructed
in terms of the metric and the curvature tensor. In the case of
four dimensions the ultraviolet divergences arise as $s \to 0$
in the first three terms of the series. The corresponding
adiabatic expansion coefficients are given by,
\ba
&a_{0} = 1&\\ \nonumber
&a_{1} = {1\over 6}{R}&\\ \nonumber
&a_{2} = {1\over 180}{R^{abcd}R_{abcd}} - {1\over 180}{R^{ab}R_{ab}}
+ {1\over 30}{{\bigtriangledown}^{a}{\bigtriangledown}_{a}R}
+ {1\over 72}R^{2}&
\ea 
The effective action may be regulated using many different
methods. It is usually found [8, 13] that the term 
linear in the curvature,$R$,
is quadratically
divergent in the corresponding regularizing parameter and
renormalizes the bare gravitational constant, $i.e$,
we have,
\be
{1\over G_{R}} = {1\over G} + a
\ee
here ${G_{R}}$ is the renormalized gravitational
constant and $a$ is the quantum correction arising
from the scalar field sector.

A related issue is the entropy of matter fields
defined on the black hole backgrounds and in thermal
equilibrium with the black hole at a temperature
corresponding to that of the black hole event horizon.
When one tries to find the degeneracy of the field
modes, one finds an alarming divergence associated
with the infinite blue shift at the event horizon.
To resolve the problem 't Hooft proposed a model
in which only low energy quantum fluctuations of
the fields are taken into account [3]. He assumed
that the usual quantum field theoretic description
of the matter fields are valid up to some point
close to the event horizon: $r_{1} = r_{+} + \epsilon,
~\epsilon \ge 0$. Correspondingly we put the following
boundary condition on the scalar field wave function,
\be
\psi(r,\theta,\phi,t) = 0 ~ if ~ r \leq r_{1}.
\ee
For asymptotically flat space time we also need an
infrared cutoff in the form of box with a large radius $L$:
\be
\psi(r,\theta,\phi,t) = 0 ~ if ~ r \geq L.
\ee
The radial degeneracy factor is obtained through
a semiclassical quantization condition using the
WKB approximation. The entanglement entropy is
obtained through explicit counting of states.
The entanglement entropy of a minimally coupled
scalar field defined in asymptotically
flat  nonextreme
Reissner-Nordstrom black hole
was obtained in [5] and the ultraviolet divergent part 
to the leading order is given by,
\be
S_m = {{r_{+}}^2\over{90h^2}} 
\ee
Here $h$ is a covariant cutoff parameter and in terms
of the coordinate cutoff parameter, $\epsilon$, it is given by,
\be
h = {\int_{r_+}^{r_1}}{dr\over \sqrt{V(r)}}
\ee
The ultraviolet divergent part of the matter field
entropy and the standard Bekenstein-Hawking entropy,
${S_{BH}} = {A\over(4G)}$, where $A$ is the area of the
black hole event horizon are related in a simple
way. If we add these two entropies, we have,
\be
S_{BH} + S_{m} = {A\over 4}{\big({1\over G} + 
{1\over 360\pi h^2}\big)}
\ee
If we compare equ.(12) with equ.(7) we find that the entanglement
entropy of the scalar field is in close analogy with the quantum
correction of the scalar field to the bare gravitational constant    
arising from the one loop effective action (4). 
Similar observations in the case of Schwarzschild metric
led Susskind and Uglum to put forward, for the case of 
canonical quantum gravity coupled to matter fields,
the following conjecture:
 
{\it The expression $S_{BH} = {A\over 4 \pi}$ for the 
Bekenststein-Hawking entropy of the fields propagating 
outside a black hole is a general result, but the gravitational
coupling arising is the renormalized gravitational coupling $G_{R}$
given by equ.(7). \it}

The brick wall model was used by G 't Hooft in a different
context[3]. He wanted to explain the black hole
entropy in terms of matter fields living outside the
horizon and in thermal equilibrium with
the black bole. If we equate $S_{m}$ with
the Bekenstein-Hawking value $S_{BH}$, the cut-off
parameter is found out to be of the order of
the Planck length and it is given by
$h = \sqrt{G\over{90\pi}}$.

An exceptional situation arises when one considers the case
of the extremal Reissner-Nordstrom black hole in asymptotically
flat space time. In this case the gravitational entropy and
the temperature of the event horizon vanishes [5]. The 
entanglement entropy of a minimally coupled massive scalar
field in this case is given by [5],
\be
S_m \sim {{8\pi^3}\over{135}}{({r_+}/\beta)^3}{exp(3\lambda/{r_+})}
\ee 
where $\lambda \to \infty$ is a large cutoff parameter. The 
 exponential divergence can be traced through the fact that
the horizon is actually infinite distance away along any constant
time hypersurface. 
If we take the standard value
of the event horizon temperature, 
which is zero, the matter field entropy
vanishes and we do not face any problem. In the Euclidean
sector the horizon is infinitely far away along all directions.
This means that the Euclidean solution can be identified with
any temperature $\beta$ and the black hole can be in 
equilibrium with thermal radiation at any temperature.
However with $\beta$ arbitrary the expression (13) is exponentially
divergent and with no corresponding term from the gravitational
sector, it is not obvious how to interpret $S_m$ in terms
of quantum corrections to the renormalization of the gravitational
constant $G$.

In this context it is interesting to study the entanglement
entropy of a minimally coupled scalar field in cold Reissner-Nordstrom
de Sitter background where an extremal black hole is 
immersed in the surrounding Hawking radiation coming from
the cosmological horizon with a finite temperature.

\section{{Calculation of entanglement entropy}}

The Reissner-Nordstrom de Sitter metric is given by,
\be
ds^2 = -V(r)dt^2 + {dr^2 \over V(r)} + {r^2}(d{\theta}^2 + {\sin}^2{\theta}
d{\phi}^2)
\ee
where
\be
V(r) = 1 - {2M\over r} + {Q^2 \over r^2} - {1\over 3}{\Lambda}r^2 
\ee
for electrically charged solution the gauge field is,
\be
F = -{Q\over r^2}dt\wedge dr
\ee
This solution has three independent parameters,the 'mass' $M$,
charge $Q$, and cosmological constant $\Lambda$, all of which
are positive. There are four roots of $V(r)$, which we designate
by $r_1,r_2,r_3,r_4$ in ascending order. In the Lorentzian section,
$0 \leq r < \infty$. The first root is negative and has no physical
significance. The second root $r_2$ is the inner (Cauchy) black hole
horizon, $r = r_3$ is the outer (Killing) horizon, and $r = r_4$
is the cosmological (acceleration) horizon.
The Killing vector $K = {{\partial}\over{\partial t}}$ is uniquely
defined by the conditions that it be null on both the Killing
horizon and the cosmological horizons and that its magnitude
should tend ${{\big({\Lambda\over 3}\big)}^{1/2}}r$ as $r$ 
tends to infinity.
For the cold Reissner-Nordstrom de Sitter metric $V(r)$ has
a double root. In this case the Cauchy horizon and the Killing
horizon coincide, $i,e$ $r_2 = r_3 = \rho$ and we have,
\be
V(r) = {\big({1 - {\rho\over r}}\big)}^2
{\big[{1 - {\Lambda\over3}(r^2 + 2\rho r + 3{\rho}^2)}\big]}
\ee
the critical relationships between mass, charge, event horizon
radius and the cosmological constant is,
\ba
&M = \rho{\big({1 - {2\over3}{\Lambda}{\rho}^2}\big)}&\\ \nonumber
&Q^2 = {\rho}^{2}{\big({1 - {\Lambda}{\rho}^2}\big)}&
\ea
For $\Lambda$ positive, there is a maximum allowed radius,
$\rho = {\rho}_m = {\Lambda}^{-1/2}$, at which $Q = 0$. The
resulting metric is the Nariai metric and is characterized
by,
${M = {\rho\over 3}},~~~{Q^2 = 0},~~~{\Lambda = {\rho}^{-2}}$, and
\be
V_N(r) = -{r^2\over{3{\rho^2}}}{\big({1 - {\rho\over r}}\big)^2}
{\big({1 + {2\rho\over r}}\big)}
\ee
For $0 < \rho < \rho_m = \Lambda^{-{1\over2}}$, there is
an extra positive root $b$ to $V(r)$, given by,
\be
b = \sqrt{(3/\Lambda) - 2 \rho^2} - \rho
\ee
For the range $0 < {\rho}^2 < {2\over\Lambda}$, the extra
horizon at $b$ is outside the cold horizon at $\rho$. We
have,
\be
{M = {{\rho{(b + \rho)}^2}\over{b^2 + 2\rho b + 3{\rho}^2}}},
~~~~{Q^2 = {{b{{\rho}^2}{(b + 2\rho)}}\over{b^2 + 2\rho b + 3{\rho}^2}}},
~~~~{\Lambda = {3\over{b^2 + 2\rho b + 3{\rho}^2}}}
\ee
the metric function becomes,
\be
V(r) = {{r^2}\over{b^2 + 2\rho b + 3{\rho}^2}}
{\big({1 - {\rho\over r}}\big)^2}
{\big({{b\over r} - 1}\big)}{\big({1 + {{2\rho + b}\over r}}\big)}
\ee
the Hawking temperature, which is given by the surface gravity
at the cosmological horizon at $b$, is:
\be
T_b =  {{b}\over{2\pi{b^2 + 2\rho b + 3{\rho}^2}}}
{\big({1 - {\rho\over b}}\big)^2}{\big({1 + {{\rho}\over b}}\big)}
\ee
for small $\Lambda~~(\Lambda \ll \rho^{-2}, or b \gg \rho)$,
$b$ is the outer edge of the cosmological horizon. 
In the Euclidean sector the double root in $V(r)$ implies that
the proper distance from any point between $\rho$ and $b$ to
$r = \rho$ along spacelike directions is infinite. In this
case, we may obtain a regular instanton by identifying the
imaginary time $\tau$
periodically with period $T_b = {{k_b}\over{2\pi}}$, where
${k_b}$ is the surface gravity of the cosmological event
horizon. 
It is obvious that the cold Reissner-Nordstrom de Sitter solution
represents an extremal black hole immersed in the thermal bath of
the Hawking radiation coming from the cosmological horizon. Consequently,
in the corresponding expression to equ.(13), we can take the temperature
to be that of the cosmological horizon.

The wave equation for a minimally coupled scalar field in a
curved background is
\be
{1\over\sqrt{-g}}{\partial_{\mu}}({\sqrt{-g}}{g^{\mu\nu}}{\partial_{\nu}{\psi}})
-m^{2}{\psi} = 0.
\ee
The region of physical interest is $\rho<r<b$. In this region
the space time is static and allows a global time-like
killing vector. 
Since there is no explicit time dependence in the foliation
used, we may construct stationary state solutions. 
We take the stationary state solutions to be of the
form
\be
{\psi} = K{e^{iEt}}{e^{iN{\phi}}}{P({\theta})}{R(r)}
\ee
where $N$ is an integer, $K$ a normalization constant, $E$ a real parameter
and represents the energy of the scalar field.
We have, using the separation of variables,
\be
{1\over{\sin{\theta}}}{\partial_{\theta}({\sin{\theta}}{\partial_{\theta}P})}
+ {[{l}({{l}+1}) - {N^{2}\over{\sin^{2}{\theta}}}]P} = 0,
\ee
The solution of this equation is given by the standard
associated Legendre polynomials.
For the radial part, we have,
\be
{E^2\over V(r)}R(r) + {1\over r^2}{\partial_r}[{r^2}V(r){\partial_r}R(r)]
- {\big({{l(l + 1)\over r^2} + m^2}\big)}R(r) = 0
\ee
We want to calculate the entropy of the scalar field. 
For this purpose we use the WKB approximation to the radial
differential equation to obtain the radial degeneracy factor
associated with the brick wall boundary condition. 
In this case the  blue-shift factor diverges at both the
black hole event horizon and the cosmological event
horizon. Following 't Hooft's procedure, we introduce
two brick wall cut-offs, one near the black hole event
horizon and the other near the cosmological horizon 
by setting
\ba
R(r) = &{0 ~~~ {\rm for \rm} ~~~~ r \leq {\rho + \epsilon_1}}&\\ \nonumber
     = &{0 ~~~ {\rm for \rm} ~~~~ r \geq {b - \epsilon_2}}&
\ea
where $\epsilon_1,\epsilon_2 \ll \rho$.

In the WKB approximation we have $R(r) = {\alpha(r)}{e^{iS(r)}}$,
where $\alpha(r)$ is
a slowly varying amplitude and $S(r)$ is a rapidly
varying phase factor. When substituted in equ.(11),
this gives us the following value for the $r$-dependent
wave number  
\be
{k^2}{(r,l,E)} = {1\over {V(r)}^2}
{\big[{E^2} - V(r){\big({{l(l + 1)\over r^2} + m^2}\big)}\big]}
\ee
The radial degeneracy factor is obtained through the following
semiclassical quantization condition:
\be
\pi{n_r{(l,E)}} = {{\int_{\rho + \epsilon_1}^{b - \epsilon_2}}
k(r,l,E)dr}
\ee
where it is implicitly assumed that the integration
is carried over those values of $l$ for which $k(r,l,E)$
is real.

The total number of wave solutions with energy not exceeding $E$,
$g(E)$, is then given by,
\be
g(E) = {\int}(2l + 1)dl~{\pi{n_r{(l,E)}}}
\ee
Every energy level determined as above may be occupied with
any nonnegative number of quanta. The free energy at an
inverse temperature $\beta$ is:
\ba
\pi\beta F &=& {{\int}dg(E) ln(1 - e^{\beta E})} \\ \nonumber
          &=&{-\beta}{\int_{0}^{\infty}}{{dE}\over{(e^{\beta E} - 1)}}}
{\int_{\rho + \epsilon_1}^{b - \epsilon_2}{dr\over V(r)}}
{{\int}{(2l + 1)dl}{\sqrt{E^2 - V(r)[{m^2 + {l(l + 1)}\over r}]}}
\ea
When we perform the $l$ integration over the values for which
the square root is positive, we have,
\be
\beta F = -{2\beta \over 3\pi}
{{\int}_{0}^{\infty}}{{dE}\over{(e^{\beta E} - 1)}}
{\int_{\rho + \epsilon_1}^{b - \epsilon_2}{dr{{r^2}\over{V(r)}^2}}
{[E^2 - V(r)m^2]}}
\ee
To the leading order this gives us the following contribution
to the free energy of the scalar field:
\be
F(\beta) = -{{{2{\pi}^3}\over{15{\beta}^4}}
{{\rho^6}{(b^2 + 2b\rho + 3{\rho}^2)^2}\over {{(b - \rho)^2}{(b + 3\rho)^2}
{\epsilon_1}^3}}
-
{{{\pi}^3}\over{90{\beta}^4}}
{{b^6}{(b^2 + 2b\rho + 3{\rho}^2)^2}\over {{(b - \rho)^2}{(b + \rho)^2}
{\epsilon_2}}}}
\ee
The radial brick wall cut-off parameters as introduced in
equation (28) are non-covariant. We replace these by the covariant
cut-off parameter $h$ as introduced in equation (11).
For the case of the cosmological horizon, $r = b$, we have,
\be
h_2 = \sqrt{{{b^2}{(b^2 + 2b\rho + 3{\rho}^2)^2}}\over 
{{2(b - \rho)^2}{(b + \rho)}}}  
{\int_{b - \epsilon_2}^{b}}{dr\over{\sqrt{b - r}}}
\ee
this gives,
\be
\epsilon_{2} = {{{(b - \rho)^2}{(b + \rho)}}\over{2{b^2}{(b^2 + 2b\rho +
3{\rho}^2)}}}{h_2}^2
\ee
For the the case of the black hole Killing horizon, this issue
is non-trivial due to presence of the double root in the metric
function at $r = \rho$. In this case we have,
\be
h_1 = \sqrt{{{(b^2 + 2b\rho + 3{\rho}^2)}}\over
{{(b - \rho)}{(b + 3\rho)}}}
{\int^{\rho + \epsilon_1}_{\rho}}{rdr\over{r - \rho}}
\ee
this is undefined at the horizon. Following [5], we set the cut-off
in terms of the proper radial variable, defined by,
\be
ds = {{dr^2}\over{{V(r)}^2}}
\ee
The horizon is at $s = -\infty$. Thus the cut-off is at a large
negative distance $s = -\lambda$, where
\be
-\lambda = p\rho{[1 + {ln{(\epsilon_1 /\rho)}}]} + p\epsilon_1,
~~~p = \sqrt{{{(b^2 + 2b\rho + 3{\rho}^2)}}\over
{{(b - \rho)}{(b + 3\rho)}}}
\ee
this gives,
\be
\epsilon_1 \sim {\rho exp[{-({\lambda\over p\rho})}]} 
\ee
The entropy is given by
\be
S = {\beta^2}{dF\over d\beta}
\ee
when $\epsilon_1,\epsilon_2$ are replaced by the covariant
cut-off parameters and 
for the value of temperature as given in  expression (23),
this leads to,
\be
S \sim {{\rho^3 \over {15{b^3}}}exp{({3\lambda\over p\rho})} +
{{b^2}\over{90 h^2}}}
\ee                        
here $\lambda \to \infty$ and $h \to 0$. 

In the case of the Euclidean lukewarm solution the conical
singularities at both the outer Killing horizon and the
cosmological event horizon are removed by identifying the
imaginary time coordinate with the same period. 
In this case the black hole horizon is at thermal equilibrium
with the cosmological horizon.
This leads
to the following condition,
\be
V(a) = V(b);~~~~V'(a) = V'(b)
\ee
this leads to,
\be
{V_{lu}(r)} = {{(r - a)(b - r)}\over{{r^2}(a + b)^2}}
{[r^2 + r(a + b) -ab]}
\ee
where $a$ and $b$ are the outer black hole horizon and the
cosmolological horizon. We have the following relations:
\be
M = {ab\over(a + b)},~~~~~Q^2 = [{ab\over(a + b)}]^2,
~~~~~~\Lambda = {3\over(a + b)} 
\ee
The common temperature at $a$ and $b$ is,
\be
T = {{(b -a)}\over{2\pi(a + b)^2}}
\ee
The entanglement entropy of a minimally coupled scalar
field, to the leading order, is given by,
\be
S = {{a^2}\over{1440{h_1}^2}} + {{b^2}\over{1440{h_2}^2}}
\ee
where ${h_1},{h_2} \to 0$, are the two covariant cut-off parameters
associated with the black hole horizon and the cosmological horizon
respectively.

\section{{Discussion}}

The Bekenstein-Hawking
entropy for the cold Reissner-Nordstrom de Sitter black hole
was obtained by Mann and Ross [11]. It is given by,
\be
S_{BH} = {{\pi}b^2} ~= A/4
\ee
The total entropy is then given by:
\ba
S = &S_{BH} + S_{m}&\\ \nonumber
  = &{{\pi}b^2}{\big({{1\over G} + {1\over{90{\pi}{h_2^2}}}}\big)} +
     {{{\rho^3}\over{15b^3}}{exp(3\lambda)}}&\\ \nonumber
  = &{A\over{4{G_R}}} + {{\rho^3}\over{15b^3}}{exp(3\lambda)}& 
\ea
where $G_R$ is the renormalized gravitational constant. As
 mentioned earlier cold Reissner-Nordstrom de Sitter
black hole represents an extremal black hole
immersed in the de Sitter space. Here, in the corresponding
expression to equ.(13) we have a natural choice for the value
of temperature, the temperature of the cosmological horizon.
In this case we find that the gravitational constant
is renormalized through the ultraviolet divergent part
of the scalar field entropy associated with the cosmological
horizon. The divergent contribution associated with the 
black hole horizon, apart from the exponential factor,
behaves like the infrared divergent part arising in the asymptotically
flat Reissner-Nordstrom black hole. 
The black hole horizon plays the role of a boundary
at the internal infinity.    

In the case of 
the lukewarm solution the Bekenstein-Hawking entropy is
given by,
\be
S_{BH} = {{\pi}a^2} + {{\pi}b^2} ~~= (A_B + A_C)/4
\ee
here $A_B$ and $A_C$ are the black hole horizon and the cosmological
horizon surface area. The total entropy is given by,
\ba
S = &S_{BH} + S_m&\\ \nonumber
  = &{{\pi}a^2}{\big({{1\over G} + {1\over{1440\pi h_1^2}}}\big)} +
     {{\pi}b^2}{\big({{1\over G} + {1\over{1440\pi h_1^2}}}\big)}&
\ea
Now, without any lose of generality, we can identify $h_1~and~h_2$.
In that case we have,
\be
S = (A_B + A_C)/{4G_R}
\ee
$i.e$, the ultraviolet divergences of the scalar field entanglement
entropy 
associated with the black hole
event horizon and the cosmological event horizon 
gives quantum corrections to the renormalized gravitational
constant $G_R$.
 
As discussed earlier we can fix the value of the covariant
cut-off parameter $'h'$ by equating the entanglement
entropy to the gravitational entropy. In the case of the
cold Reissner-Nordstrom de Sitter black hole, we can equate
the ultraviolet divergent part associated with the cosmological
horizon of the scalar field entanglement
entropy to the corresponding gravitational entropy. The cut-off
parameter "h" in this case is given by $h = \sqrt{G\over{90\pi}}$.
In the case of the lukewarm solution the cut-off is given by
$h = \sqrt{G\over{1440\pi}}$.
In the units chosen both the cut-off are of the order of the Planck
length although their values are different. This is expected
because  the
cold Reissner-Nordstrom de Sitter black hole and the
lukewarm Reissner-Nordstrom de Sitter black hole
are geometrically two different spacetime.  

We conclude with a brief comment on the Schwarzschild-de Sitter 
metric. In this case the black hole horizon radius is very
small compared to that of the cosmological horizon radius.
Consequently, the black hole Hawking radiation is at higher
temperature than that of the Hawking radiation coming from
the cosmological horizon and it is not obvious how to define
the temperature of the matter field between the two horizon.
Similar situation also arises in the case Reissner-Nordstrom
de Sitter black hole when $M \not= Q$. The entanglement
entropy of a scalar field was carried out in [15] in the
limit $r_4 \gg r_3$. In this case the temperature at which
the black hole radiates is much larger than that of the cosmological
horizon, and, the temperature of the matter field was taken
to be that of the outer Killing horizon.
However, in the semicassical appriximation, $G\Lambda \ll 1$,
the  Schwarzschild-de Sitter 
black hole evaporates away on a time scale much shorter
than that required for another nucleation [9] and we can
preferably define a test matter field at the temperature of the 
cosmological horizon. This issue is currently under study.   

\section{{Acknowledgments}}
It is a great pleasure to the author to thank
Prof. P. Mitra for many helpful
discussions.

\end{document}